\documentclass[lettersize,journal]{IEEEtran}
\usepackage{amsmath,amsfonts}

\usepackage{algorithm}
\usepackage{algpseudocode}

\usepackage{array}
\usepackage[caption=false,font=normalsize,labelfont=sf,textfont=sf]{subfig}
\usepackage{textcomp}
\usepackage{stfloats}
\usepackage{url}
\usepackage{xcolor}
\usepackage{verbatim}
\usepackage{graphicx}
\usepackage{cite}
\usepackage{xcolor}
\usepackage{tabularx,booktabs,textcomp}
\usepackage{amsthm} 

\usepackage{multirow}
\usepackage{graphicx}

\usepackage{amsmath,amssymb,amsthm}

\begin{document}
\title{Detection of High-Impedance Low-Current Arc Faults at Electrical Substations}

\author{{K. Victor Sam Moses Babu,~\IEEEmembership{Member,~IEEE}, Divyanshi Dwivedi,~\IEEEmembership{Student Member,~IEEE}, Marcelo Esteban Valdes,~\IEEEmembership{Fellow,~IEEE}, Pratyush Chakraborty,~\IEEEmembership{Senior Member,~IEEE}, Prasanta Kumar Panigrahi, Mayukha Pal,~\IEEEmembership{Senior Member,~IEEE}}

\thanks{(Corresponding author: Mayukha Pal)}

\thanks{K. Victor Sam Moses Babu is a Data Science Research Intern at ABB Ability Innovation Center, Hyderabad 500084, India and also a Research Scholar at the Department of Electrical and Electronics Engineering, BITS Pilani Hyderabad Campus, Hyderabad 500078, IN.}
\thanks{Divyanshi Dwivedi is with ABB Ability Innovation Center, Hyderabad 500084, IN, working as R\&D Engineer.}
\thanks{Marcelo Esteban Valdes is an IEEE Fellow working with ABB Inc., Cary, USA as Manager – Applications \& Standards.}
\thanks{Pratyush Chakraborty is an Asst. Professor with the Department of Electrical and Electronics Engineering, BITS Pilani Hyderabad Campus, Hyderabad 500078, IN.}
\thanks{Prasanta Kumar Panigrahi is  the Director and Founding Professor with the Centre for Quantum Science and Technology, Siksha 'O' Anusandhan University, Bhubaneswar, 751030, Odisha, India, IN.}
\thanks{Mayukha Pal is with ABB Electrification, Ability Innovation Center, Hyderabad 500084, IN, working as Global R\&D Leader – Cloud \& Advanced Analytics (e-mail: mayukha.pal@in.abb.com).}
}

\maketitle

\begin{abstract}

Arcing faults in low voltage (LV) distribution systems associated with arc- flash risk and potentially significant equipment damage are notoriously difficult to detect under some conditions. Especially so when attempting to detect using sensing at the line, high voltage side of a substation transformer. This paper presents an analytics-based physics-aware approach to detect high-impedance, low-current arcing faults from the primary side of the substation transformer at current thresholds, below normal operating events, along with transformer inrush currents. The proposed methodology leverages the Hankel Alternative View Of Koopman Operator approach to differentiate arcing faults from standard operations, while the Series2Graph method is employed to identify the time of fault occurrence and duration. Unlike prior studies that detect such faults at the device or secondary transformer side, this work demonstrates successful fault detection at the primary side of the distribution substation transformer for faults occurring on the secondary side. The approach addresses the practical challenges of differentiating primary side expected and acceptable transients from similar magnitude LV arcing fault currents that may occur on the secondary side. The results demonstrate the efficacy of the proposed method in accurately identifying fault occurrence and duration, minimizing the risk of false positives during similar characteristic events, thus improving the reliability and operational efficiency of power distribution systems. This approach can benefit both traditional and smart power grids that employ similar transformer configurations.

\end{abstract}

\begin{IEEEkeywords}
Power Distribution Systems; High Impedance Faults; Arc-Flash, Arcing Current Detection; Hankel Alternative View of Koopman Operator; Graph Theory.
\end{IEEEkeywords}

\section{Introduction}
\label{section:Introduction}
\subsection{Background and Motivation}

High-impedance faults (HIFs) in medium-voltage distribution networks pose significant operational and safety risks. In medium voltage (MV) distribution systems, these faults commonly occur when an energized conductor comes into contact with high-impedance surfaces such as asphalt, sand, or vegetation \cite{arcdd1}. In LV systems they may occur for a broad range of reasons the most concerning of which is human error associated with Arc Flash incidents. The resulting fault current is typically too low to be detected by conventional overcurrent protection devices such as circuit breakers and relays, which are designed to detect higher-magnitude faults. HIFs are characterized by their random, asymmetric, and nonlinear behavior, with current magnitude that may be below expected transient currents or even load current, hence complicating the desire for fast detection detection and fast protection \cite{arcdd2}. Although the current associated with HIFs may not be large enough to directly damage power system equipment, undetected HIFs pose severe risks to personnel, the environment, and the electrical network. Fires, including wildfires, electric shocks, and other hazards can result from undetected HIFs, making the early and accurate detection of these faults critical to maintaining the safety and reliability of distribution systems \cite{arcdd3,GHADERI2017376}.

Despite the serious risks posed by HIFs, their detection remains a formidable challenge. Studies show that in medium-voltage distribution systems (typically 4 kV to 34.5 kV), conventional protection systems clear less than 18\% of staged HIF events, exposing these distribution systems to unmitigated hazards  \cite{Rai2022, arc1}. The unique characteristics of HIFs, particularly the zero-crossing phenomenon and nonlinear distortions at the zero-crossing points, offer some potential for detection \cite{arcdd1, arc2}. It's important to note that these detectable characteristics are often more pronounced in LV systems, where the lower driving voltage makes zero-crossing distortions more evident and the nonlinear aspects more significant relative to the fundamental frequency. This voltage-dependent variation in HIF characteristics presents both opportunities and challenges for detection methods across different voltage levels. However, these phenomena, while distinguishing HIFs from other disturbances, may not produce fault currents large enough for traditional overcurrent protection devices to respond effectively.

The characteristics of HIFs vary significantly between medium voltage and low voltage systems. In MV systems, the high impedance of the surfaces involved plays a significant role in limiting fault current. However, in LV systems, it is primarily the arc impedance that dominates, typically constituting around 50\% or more of the total fault impedance. These system-specific characteristics render many conventional protection methods insufficient, allowing HIFs to remain undetected in 5-10\% of fault cases \cite{arc2, arc3}. This variation in fault impedance sources between MV and LV systems presents unique challenges for HIF detection across different voltage levels. The failure to detect HIFs reliably leads to significant safety hazards, as these faults contribute to electric shocks, fires, and premature equipment aging, ultimately resulting in higher operational costs and increased maintenance.

Research on HIF detection over the past several decades has explored a variety of methods, broadly categorized into threshold-based and artificial intelligence (AI)-based techniques. Threshold-based methods involve extracting key features from the fault signals, such as waveform distortion indicators \cite{th2}, interval slopes \cite{th1}, Kullback–Leibler
divergence \cite{th4} and Pearson skewness coefficients \cite{th3}, to differentiate HIFs from other disturbances. While these methods are relatively straightforward to implement, the inherent variability in fault characteristics, such as neutral ground modes, fault resistance, and environmental factors, makes it challenging to define a constant threshold suitable for all scenarios. This variability limits the practical applicability of threshold-based methods.

AI-based approaches, on the other hand, leverage machine learning algorithms to automate feature extraction and fault classification. Methods using support vector machines (SVM) \cite{ai1,Ahmadi2022}, decision trees \cite{ai2}, artificial neural networks (ANN) \cite{ai6,ai7}, probabilistic neural network \cite{pnn1}, evolving neural network (ENN) \cite{ai5} and convolutional neural networks (CNN) \cite{ai4} have demonstrated promise in detecting HIFs. However, these techniques are highly dependent on the quality and volume of training data, and the lack of sufficient field data can impair their effectiveness. Moreover, AI models are often criticized for their lack of interpretability, a critical shortcoming in industrial and engineering applications where transparent decision-making is essential.

In addition to AI methods, time-frequency analysis (TFA) has been widely adopted to detect HIFs by considering signal behavior across both time and frequency domains. Techniques like short-time fourier transform (STFT) \cite{ta1}, Gabor-Wigner transforms \cite{ta2}, and wavelet-based approaches \cite{ta3,ta4} have been used to extract key signal features that indicate the presence of HIFs. However, TFA-based techniques often require significant computational resources, limiting their applicability in real-time protection schemes. Furthermore, the sensitivity of these techniques to sudden changes in input signals, noise, and harmonics complicates their deployment in operational environments. Many harmonic-based methods, for instance, are prone to false detections caused by events like load switching and transformer energization, further diminishing their reliability.

Given these challenges, there is a pressing need for a more robust, accurate, and computationally efficient method to detect HIFs, especially in the context of smart grid operations. The modern grid relies heavily on real-time data for monitoring, control, and decision-making. The occurrence of undetected HIFs introduces erroneous data, which can lead to incorrect decisions and missed opportunities for preventive maintenance, ultimately compromising grid resilience. To address these issues, this paper proposes a novel physics-aware, data-driven method for detecting high-impedance, low-current arcing faults at the primary side of distribution substation transformers. Unlike previous approaches that focus on fault detection at the device or secondary transformer level, this method leverages the Hankel alternative view of Koopman operator \cite{HAVOK_chaos,dwivedi2023dynamopmu} for distinguishing HIFs from normal operating events based on system dynamics.

Additionally, the fault occurrence and duration are extracted using the Series2Graph method \cite{Series2GraphPaper}, offering a detailed temporal characterization of the fault. By focusing on the primary side of the transformer, the proposed method addresses the practical challenges of high harmonic content and the need for lower-rated current measuring devices while still accurately identifying faults that occur on the secondary side of the system. This approach not only improves fault detection but also mitigates the risk of false positives, which are often triggered by disturbances that share similar characteristics with HIFs, such as load switching or capacitor energization.

Thus, the proposed Hankel alternative view of Koopman operator model and Series2Graph-based methodology presents a significant advancement in HIF detection, offering a reliable, computationally feasible solution that enhances the security, safety, and reliability of medium-voltage distribution systems. Through the integration of advanced signal processing and machine learning techniques, the methodology provides a new avenue for effective HIF detection, ensuring optimal grid performance in the face of unpredictable and dangerous fault conditions.

The structure of this paper is as follows: Section \ref{section:Material} outlines the methodology, detailing the event detection and prediction process. Section \ref{section:Result} presents the analysis and discussion of the results. Finally, Section \ref{section:Conclusion} provides the conclusion.

\section{Materials and Methods}
\label{section:Material}

\subsection{Mathematical Modeling of Fault and Normal Events}
This study considers three types of events in medium voltage distribution systems: high-impedance arc faults, motor starting, and load switching. To accurately represent these events in our analysis, we employ the following mathematical models:
\subsubsection{High Impedance Arc Fault Model}
HIFs are modeled as a time-varying arc in series with a constant resistance. The voltage-current relationship is described by \cite{7390312}:
\begin{equation}
v_f(t) = i_f(t) \cdot R_0 + \frac{i_f(t)}{g_{arc}(t)}
\end{equation}
where $v_f(t)$ is the fault voltage, $i_f(t)$ is the fault current, $R_0$ is the constant resistance, and $g_{arc}(t)$ is the time-varying arc conductance. The dynamic behavior of the arc conductance is modeled using the Kizilcay's model:
\begin{equation}
\frac{dg_{arc}(t)}{dt} = \frac{1}{\tau} \left( \frac{|i_f(t)|}{u_0 + r_0|i_f(t)|} - g_{arc}(t) \right)
\end{equation}
Here, $\tau$ is the arc time constant, $u_0$ is the characteristic arc voltage, and $r_0$ is the characteristic arc resistance.
\subsubsection{Motor Starting and Load Switching Models}
Both motor starting and load switching events are represented by a general R-L model:
\begin{equation}
v_f(t) = i_f(t) \cdot R + L \cdot \frac{di_f(t)}{dt}
\end{equation}
where $R$ is the constant resistance and $L$ is the constant inductance. The values of $R$ and $L$ differ for motor starting and load switching events:

For motor starting: $R$ represents the stator resistance and starting impedance, while $L$ represents the transient inductance of the motor.
For load switching: $R$ and $L$ represent the equivalent impedance of the switched load.

\subsubsection{Model Application}
These models form the basis for generating the voltage and current waveforms used in our study. By varying the parameters of these models, we can represent a wide range of HIF scenarios, motor starting conditions, and load switching events. The resulting waveforms are then used to evaluate the performance of our proposed HIF detection method. The specific implementation of these models in our simulation environment, along with the details of the test cases, are presented in Section \ref{section:Result}.

\subsection{Computation of Forcing Operator using Hankel alternative view of Koopman Operator}

Hankel alternative view of Koopman operator analysis is a technique that leverages the Koopman operator theory and dynamic mode decomposition (DMD) to analyze nonlinear dynamical systems \cite{dwivedi2023dynamopmu}. It provides a framework for identifying and characterizing low-dimensional structures within high-dimensional data, facilitating the understanding and prediction of complex behaviors. This approach is particularly useful for fault detection in complex systems by capturing the underlying dynamics and identifying deviations from normal operation.

\subsubsection{Hankel Matrix Construction}

For a given time series $\{x(t_i)\}_{i=1}^N$, we construct the Hankel matrix $\mathbf{H}$ to perform time-delay embedding. This process transforms the original time series into a higher-dimensional space, revealing the underlying manifold structure essential for capturing the nonlinearity in the dynamics. The Hankel matrix $\mathbf{H}$ is constructed as follows \cite{havok}:
\begin{equation}
    \mathbf{H} = \begin{bmatrix}
x(t_1) & x(t_2) & x(t_3) & \cdots & x(t_k) \\
x(t_2) & x(t_3) & x(t_4) & \cdots & x(t_{k+1}) \\
x(t_3) & x(t_4) & x(t_5) & \cdots & x(t_{k+2}) \\
\vdots & \vdots & \vdots & \ddots & \vdots \\
x(t_m) & x(t_{m+1}) & x(t_{m+2}) & \cdots & x(t_N)
\end{bmatrix}
\end{equation}

where $m = N - k + 1$ and $k$ is the window length. The choice of $k$ is crucial as it determines the balance between capturing sufficient dynamics and maintaining computational efficiency.

\subsubsection{Singular Value Decomposition (SVD)}

We perform Singular Value Decomposition (SVD) on the Hankel matrix $\mathbf{H}$ to obtain a low-rank approximation. SVD decomposes $\mathbf{H}$ into three matrices \cite{koopman}:
\begin{equation}
    \mathbf{H} = \mathbf{U} \mathbf{\Sigma} \mathbf{V}^T
\end{equation}

where:
\begin{itemize}
    \item $\mathbf{U}$ is an $m \times m$ orthogonal matrix whose columns are the left singular vectors.
    \item $\mathbf{\Sigma}$ is an $m \times k$ diagonal matrix containing the singular values, sorted in descending order.
    \item $\mathbf{V}$ is a $k \times k$ orthogonal matrix whose columns are the right singular vectors.
\end{itemize}
The singular values in $\mathbf{\Sigma}$ provide insight into the intrinsic dimensionality of the data, allowing us to truncate the decomposition and retain only the most significant modes.

\subsubsection{Koopman Operator and Dynamic Mode Decomposition}

The Koopman operator $\mathcal{K}$ is an infinite-dimensional linear operator that acts on observable functions of the state space of the dynamical system. Despite the nonlinearity of the original system, $\mathcal{K}$ evolves observables linearly, making it a powerful tool for analyzing complex dynamics \cite{koopman}. In practice, we approximate $\mathcal{K}$ using the DMD algorithm, which involves the following steps:
\begin{enumerate}
    \item Construct two matrices $\mathbf{X}$ and $\mathbf{Y}$ from time-shifted snapshots of the data.
\begin{equation}
       \begin{aligned}
           \mathbf{X} &= \begin{bmatrix}
               x(t_1) & x(t_2) & \cdots & x(t_{k-1})
           \end{bmatrix}, \\
           \mathbf{Y} &= \begin{bmatrix}
               x(t_2) & x(t_3) & \cdots & x(t_k)
           \end{bmatrix}
       \end{aligned}
    \end{equation}   
    \item Perform SVD on $\mathbf{X}$: $\mathbf{X} = \mathbf{U}_X \mathbf{\Sigma}_X \mathbf{V}_X^T$.
    \item Compute the low-rank approximation of the Koopman operator: $\mathbf{K} \approx \mathbf{U}_X^T \mathbf{Y} \mathbf{V}_X \mathbf{\Sigma}_X^{-1}$.
\end{enumerate}

We decompose the dynamics into Koopman modes $\{\phi_i\}$ and Koopman eigenvalues $\{\lambda_i\}$:
\begin{equation}
   \mathcal{K} \phi_i = \lambda_i \phi_i 
\end{equation}

These modes and eigenvalues provide a spectral decomposition of the dynamics, revealing coherent structures and their temporal behaviors.

\subsubsection{Low-Dimensional Representation}

In the Hankel alternative view of Koopman operator framework, we identify a low-dimensional representation of the system dynamics by focusing on the most significant modes obtained from SVD. The Hankel matrix $\mathbf{H}$ captures the time-delay embedding of the system, and the SVD provides a low-rank approximation. The Koopman modes and eigenvalues characterize the underlying dynamics \cite{koopman}.

The low-dimensional representation obtained from SVD is used to construct a linear dynamical system:
\begin{equation}
    \mathbf{z}(t+1) = \mathbf{A} \mathbf{z}(t)
\end{equation}

where $\mathbf{z}(t)$ is the state vector in the reduced space, and $\mathbf{A}$ is the system matrix. This linear system approximates the original nonlinear dynamics within the reduced subspace.

\subsubsection{Fault Detection}

To detect faults using the Hankel alternative view of Koopman operator framework, we follow these steps:
\begin{enumerate}
    \item Train the Hankel alternative view of Koopman operator model on data from normal operating conditions to establish a baseline model of the system dynamics.
    \item Monitor the system in real-time, continuously constructing the Hankel matrix and applying the trained model.
    \item Compare the observed Koopman modes and eigenvalues with the baseline. Significant deviations indicate potential faults or anomalies.
\end{enumerate}
By leveraging the low-dimensional linear representation, Hankel alternative view of Koopman operator facilitates real-time fault detection and diagnosis.

\begin{figure*}
  \centering
  \includegraphics[width=6.4in]{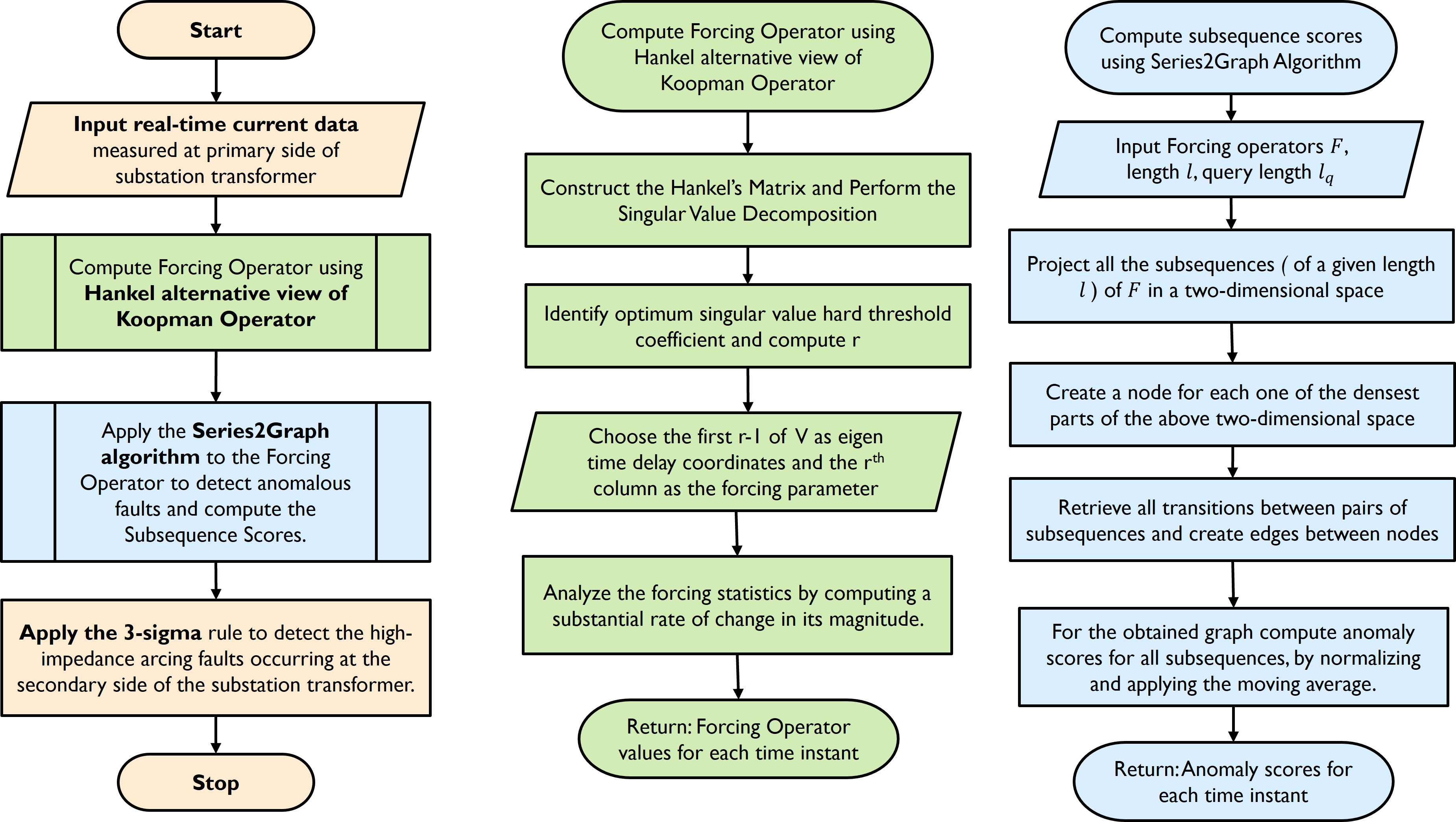}
  \caption{Flow diagram of the proposed methodology.}
  \label{fig:Method}
\end{figure*}

\subsection{Series2Graph algorithm to detect faults and compute the subsequence Scores}

Series2Graph is a graph-based approach for subsequence anomaly detection in time series data \cite{Series2GraphPaper}. Let the graph $G = (N, E)$, where:
\begin{itemize}
    \item $N$ is the set of nodes representing subsequences of the time series.
    \item $E$ is the set of directed edges representing transitions between subsequences.
\end{itemize}

The graph $G$ is defined as:
\begin{equation}
G = (N, E)
\end{equation}

\subsubsection{Node Creation}
We define nodes based on subsequences of a time series $T$. Let $T = \{T_1, T_2, \dots, T_n\}$ be a time series of length $n$, where each data point $T_i \in \mathbb{R}$ is real-valued. A subsequence of length $\ell$, starting at position $i$, is given by:
\begin{equation}
T_{i, \ell} = \{T_i, T_{i+1}, \dots, T_{i+\ell-1}\}
\end{equation}
Each subsequence $T_{i, \ell}$ is mapped to a node $N_i$ in the graph, and the set of nodes is:
\begin{equation}
N = \{N_1, N_2, \dots, N_m\}
\end{equation}
where $m = n - \ell + 1$.

\subsubsection{Edge Creation}
Edges are created between nodes if one subsequence follows another in the original time series. Let an edge $(N_i, N_j)$ represent a transition from subsequence $T_{i, \ell}$ to $T_{j, \ell}$.

The weight of an edge $w(e)$, for $e = (N_i, N_j)$, is defined as the number of transitions from subsequence $T_{i, \ell}$ to $T_{j, \ell}$:
\begin{equation}
w(e) = \text{Number of transitions from } N_i \text{ to } N_j
\end{equation}
Thus, the set of edges is:
\begin{equation}
E = \{ (N_i, N_j) \mid \text{transitions between subsequences} \}
\end{equation}

\subsubsection{Path Normality Score}
The normality score of a subsequence is determined by examining the transitions between nodes in the graph. Given a subsequence of length $\ell_q$, the path through the graph $P_{\text{th}}$ consists of the nodes that represent the subsequence.

Let the path corresponding to subsequence $T_{i, \ell_q}$ be represented by $P_{\text{th}} = \{N_i, N_{i+1}, \dots, N_{i+\ell_q-1}\}$. The normality score $\text{Norm}(P_{\text{th}})$ is calculated as:
\begin{equation}
\text{Norm}(P_{\text{th}}) = \frac{1}{\ell_q} \sum_{j=i}^{i+\ell_q-1} \frac{w(N_j, N_{j+1})}{\deg(N_j) - 1}
\end{equation}
where $w(N_j, N_{j+1})$ is the weight of the edge between nodes $N_j$ and $N_{j+1}$, and $\deg(N_j)$ is the degree of node $N_j$.

\subsubsection{Fault Detection: Normal vs Anomalous Behavior}
To distinguish between normal and anomalous subsequences, a threshold $\theta$ is used. If the normality score of a subsequence is greater than or equal to the threshold, the subsequence is considered normal \cite{Series2GraphPaper}:
\begin{equation}
\text{Norm}(P_{\text{th}}) \geq \theta
\end{equation}
Anomalous subsequences, indicating faults, satisfy the condition:
\begin{equation}
\text{Norm}(P_{\text{th}}) < \theta
\end{equation}

\subsubsection{Anomaly Subgraph}
The anomaly subgraph $G_{\alpha}$ contains the nodes and edges corresponding to the anomalous subsequences. Formally, the anomaly subgraph is:
\begin{equation}
G_{\alpha} = (N_{\alpha}, E_{\alpha})
\end{equation}
where:
\begin{itemize}
    \item $N_{\alpha} \subset N$ is the set of nodes representing anomalous subsequences.
    \item $E_{\alpha} \subset E$ is the set of edges between the anomalous nodes.
\end{itemize}

\begin{algorithm}
\caption{Series2Graph for Fault Detection and Subsequence Scoring}
\begin{algorithmic}
    \State \textbf{Input:} Time series $T = \{T_1, T_2, \dots, T_n\}$, subsequence length $\ell$, query length $\ell_q$, threshold $\theta$
    \State \textbf{Output:} Anomaly subgraph $G_{\alpha} = (N_{\alpha}, E_{\alpha})$
    \State \textbf{Step 1: Extract subsequences and create nodes}
    \For{$i = 1$ to $n - \ell + 1$}
        \State Extract subsequence $T_{i, \ell} = \{T_i, \dots, T_{i+\ell-1}\}$ and create node $N_i$
    \EndFor
    \State \textbf{Step 2: Create edges}
    \For{each pair of consecutive subsequences $T_{i, \ell}$, $T_{j, \ell}$}
        \State Add edge $E = (N_i, N_j)$ and set weight $w(E)$ as number of transitions
    \EndFor
    \State \textbf{Step 3: Compute normality score}
    \For{each subsequence $T_{i, \ell_q}$}
        \State Traverse path $P_{\text{th}} = \{N_i, \dots, N_{i+\ell_q-1}\}$ and compute:
        \[
        \text{Norm}(P_{\text{th}}) = \frac{1}{\ell_q} \sum_{j=i}^{i+\ell_q-1} \frac{w(N_j, N_{j+1})}{\deg(N_j) - 1}
        \]
    \EndFor
    \State \textbf{Step 4: Fault detection}
    \For{each subsequence $T_{i, \ell_q}$}
        \If{$\text{Norm}(P_{\text{th}}) < \theta$}
            \State Mark subsequence as \textbf{anomalous} and add to $G_{\alpha}$
        \Else
            \State Mark subsequence as \textbf{normal}
        \EndIf
    \EndFor
    \State \textbf{Return} $G_{\alpha}$
\end{algorithmic}
\end{algorithm}

\subsection{Proposed Methodology}

The proposed methodology is illustrated in Fig. \ref{fig:Method}. The methodology begins with real-time current data measured at the primary side of a substation transformer as input. This data is used to compute the Forcing Operator using the Hankel alternative view of the Koopman Operator. A Hankel matrix is constructed by stacking the input time series data advanced by one measurement at each level. Singular Value Decomposition is then performed on this Hankel matrix to obtain left and right singular vectors. An optimum singular value hard threshold coefficient is identified, where the first $r-1$ columns of the right singular vector matrix represent eigen time delay coordinates, and the $r$-th column is the corresponding forcing statistic. he Series2Graph algorithm is then applied to the Forcing Operator to detect anomalous faults and compute Subsequence Scores. This algorithm projects subsequences of the Forcing Operator into a two-dimensional space, creates nodes for dense regions, and establishes edges between nodes based on transitions between subsequences. Anomaly scores are computed for all subsequences by normalizing and applying a moving average. Finally, the 3-sigma rule is applied to these scores to detect high-impedance arcing faults occurring at the secondary side of the substation transformer. This unsupervised method is domain-agnostic and can identify anomalies of varying lengths without requiring labeled instances or anomaly-free data.

\section{Results and Discussion}
\label{section:Result}

\begin{figure}[htbp]
  \centering
  \includegraphics[width=3.5in]{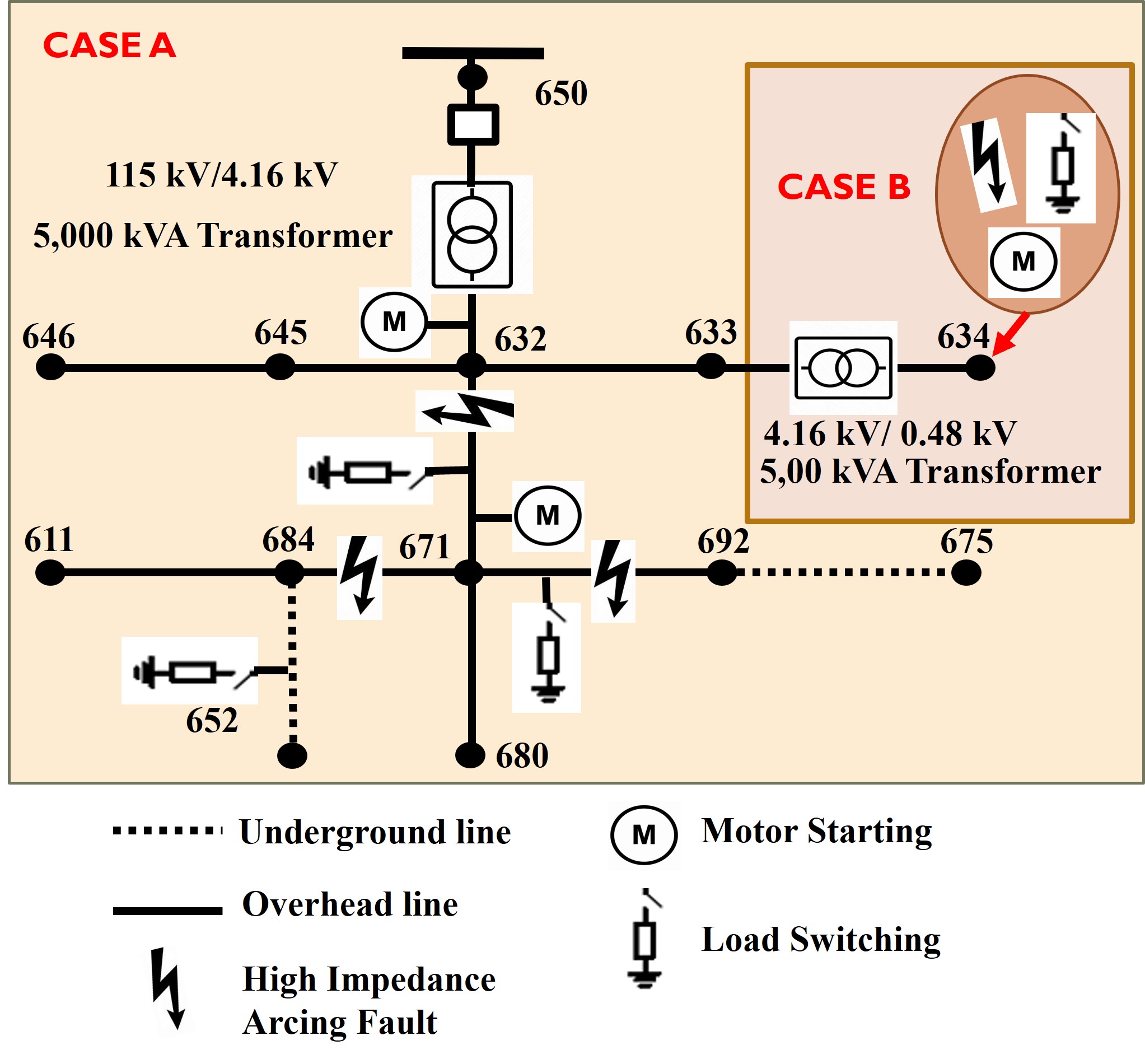}
  \caption{Modified IEEE 13-bus system simulation model.}
  \label{fig:simulation_model}
\end{figure}

In the simulation setup, illustrated in Figure~\ref{fig:simulation_model}, we modified the IEEE 13-bus system to create a comprehensive test environment for our HIF detection method. The system is divided into two main cases: Case A, which encompasses the entire distribution network with a primary substation transformer (115 kV/4.16 kV, 5,000 kVA), and Case B, which focuses on a secondary distribution transformer (4.16 kV/0.48 kV, 500 kVA). To create a more challenging and realistic scenario, we implemented a detailed model of the substation transformer that captures various transient behaviors. This model incorporates key transformer characteristics such as core saturation, magnetizing inrush currents, and harmonic distortions. By including these transient effects, we simulate the complex electromagnetic phenomena that occur during transformer energization and fault conditions. These transients can potentially mask or mimic HIF characteristics, making detection more difficult. The transformer model allows for the simulation of inrush currents during energization, which can be several times the rated current and contain significant harmonic content. Additionally, the model accounts for core saturation effects, which can introduce further nonlinearities in the transformer's response to system disturbances. By incorporating these detailed transient characteristics, our simulation creates a more realistic representation of the challenges faced in real-world HIF detection scenarios. This approach enables us to thoroughly evaluate the robustness and effectiveness of our proposed HIF detection method under conditions that closely resemble actual distribution system operations.

\begin{figure}[htbp]
  \centering
  \includegraphics[width=3.5in]{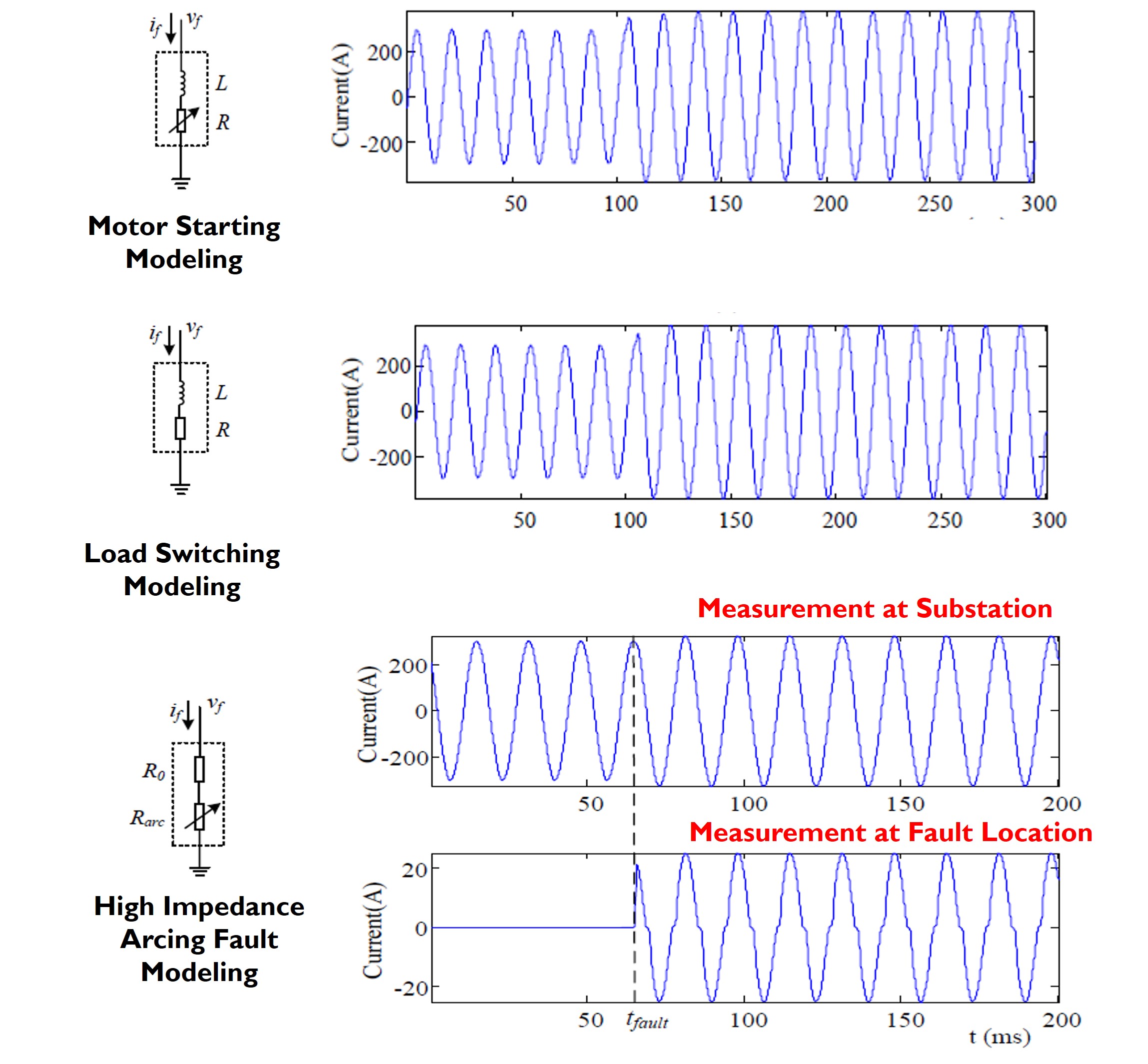}
  \caption{Current waveforms for various events in the distribution system.}
  \label{fig:waveforms}
\end{figure}

Figure~\ref{fig:waveforms} displays the current waveforms for different events: motor starting, load switching, and high impedance arcing faults. These waveforms illustrate the system's response to each event type, with measurements taken at both the substation and fault locations. It can be observed that motor starting and load switching would appear similar to the the HIF at the substation secondary side of the transformer. Tables~\ref{tab:case_a} and~\ref{tab:case_b} detail the event sequences for Case A and Case B, respectively. Case A presents a more complex scenario with multiple occurrence of events across the distribution system and the data is observed at the primary substation transformer, while Case B simulates a simplified event sequence at one node of the secondary side of the secondary substation transformer.

\begin{table}[htbp]
\centering
\caption{Event sequence for Case A}
\begin{tabular}{ll}
\toprule
Events                          & Time Occurrence  \\
\midrule
Load Switching                  & 1.1 sec, 1.3 sec, 1.5 sec \\
Motor Starting                  & 1.0 sec, 1.7 sec          \\
High Impedance Low Arcing Fault & 1.2 sec, 1.4 sec, 1.6 sec \\
\bottomrule
\end{tabular}
\label{tab:case_a}
\end{table}

\begin{table}[htbp]
\centering
\caption{Event sequence for Case B}
\begin{tabular}{ll}
\toprule
Events   & Time Instant  \\
\midrule
Load Switching                  & 1.2 sec        \\
Motor Starting                  & 1.6 sec        \\
High Impedance Low Arcing Fault & 1.4 sec       \\
\bottomrule
\end{tabular}
\label{tab:case_b}
\end{table}

\begin{figure}[htbp]
  \centering
  \includegraphics[width=3.5in]{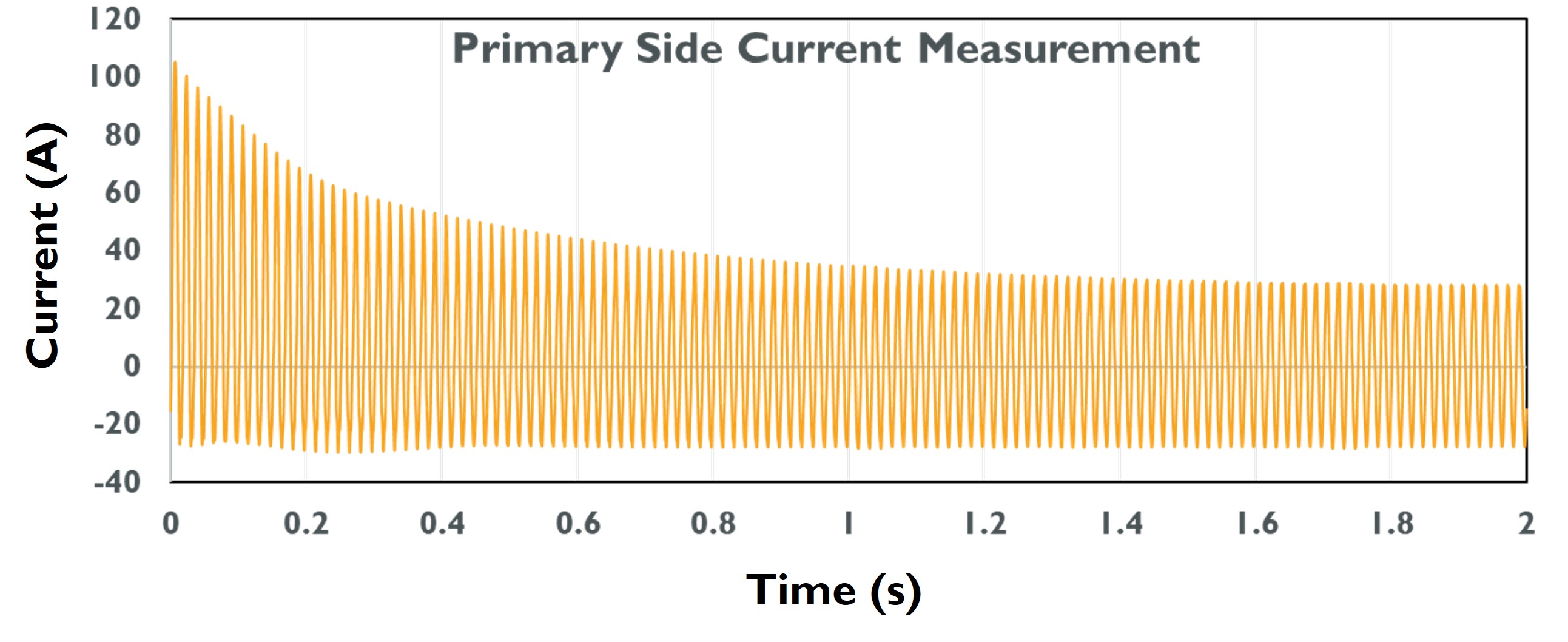}
  \caption{Primary side current measurement for Case A.}
  \label{fig:current_A}
\end{figure}

\begin{figure}[htbp]
  \centering
  \includegraphics[width=3.5in]{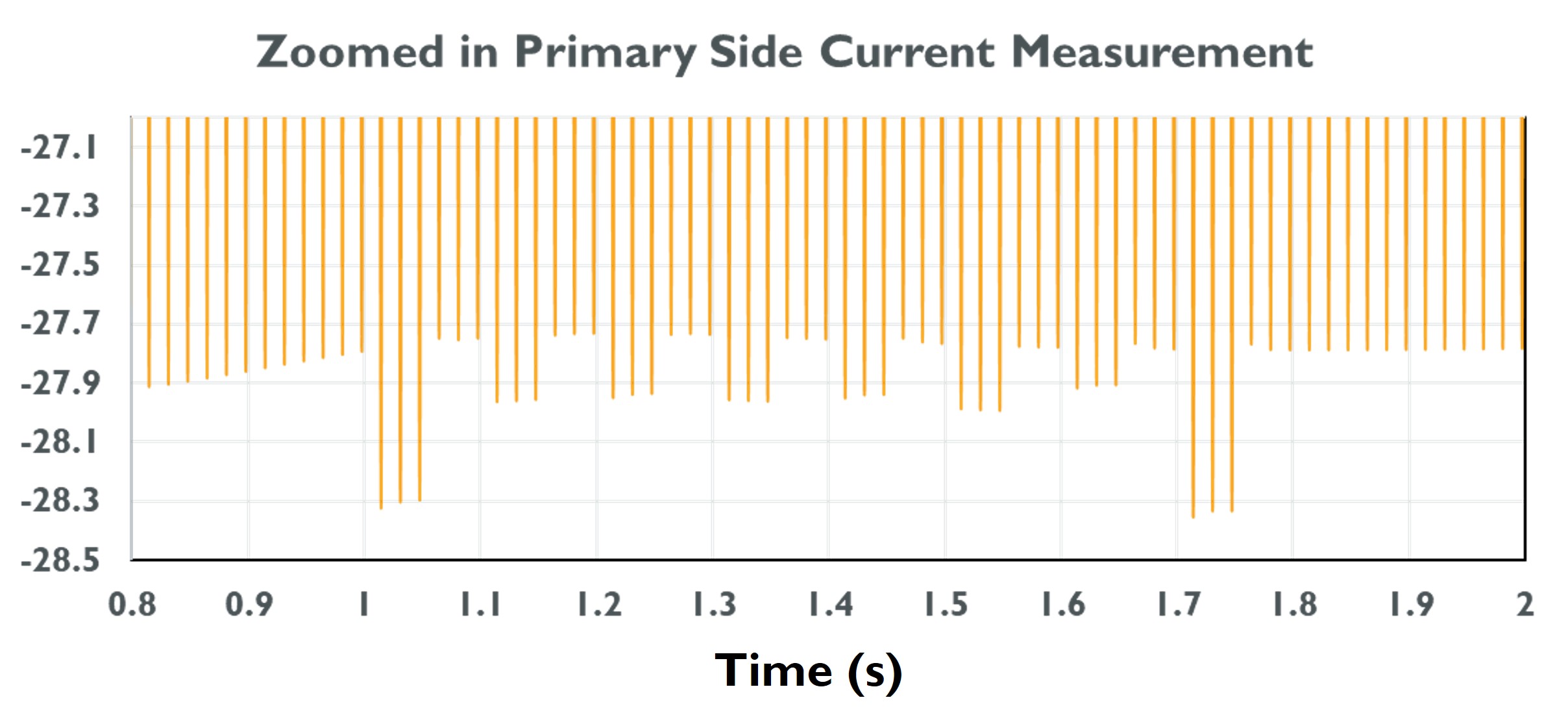}
  \caption{Zoomed view of primary side current measurement for Case A.}
  \label{fig:zoomed_current_A}
\end{figure}

The simulation results for Case A demonstrate the effectiveness of our proposed method in detecting high-impedance arcing faults at the primary side of the primary substation transformer. Figures~\ref{fig:current_A} and~\ref{fig:zoomed_current_A} show the primary side current measurements, illustrating the significant challenge in identifying faults directly from these measurements. The current waveform in Figure~\ref{fig:current_A} displays a complex pattern with varying amplitudes over the 2-second simulation period, which could mask the subtle changes caused by high-impedance faults. Figure~\ref{fig:zoomed_current_A} provides a zoomed view of the current measurement, revealing small distortions and asymmetries that could be indicative of fault conditions but are not easily distinguishable from normal operating variations.

\begin{figure}[htbp]
  \centering
  \includegraphics[width=3.5in]{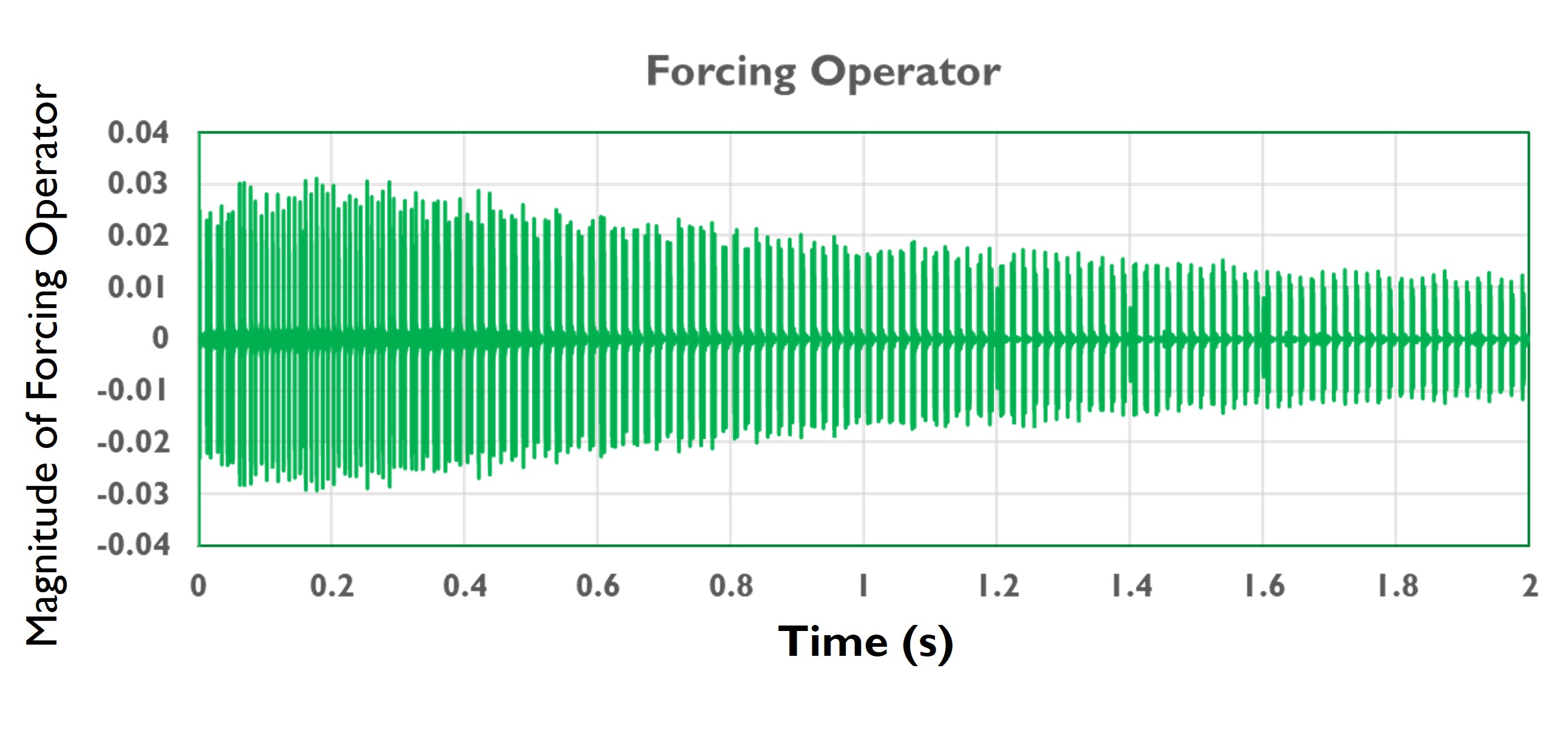}
  \caption{Forcing operator magnitude for Case A.}
  \label{fig:forcing_A}
\end{figure}

\begin{figure}[htbp]
  \centering
  \includegraphics[width=3.5in]{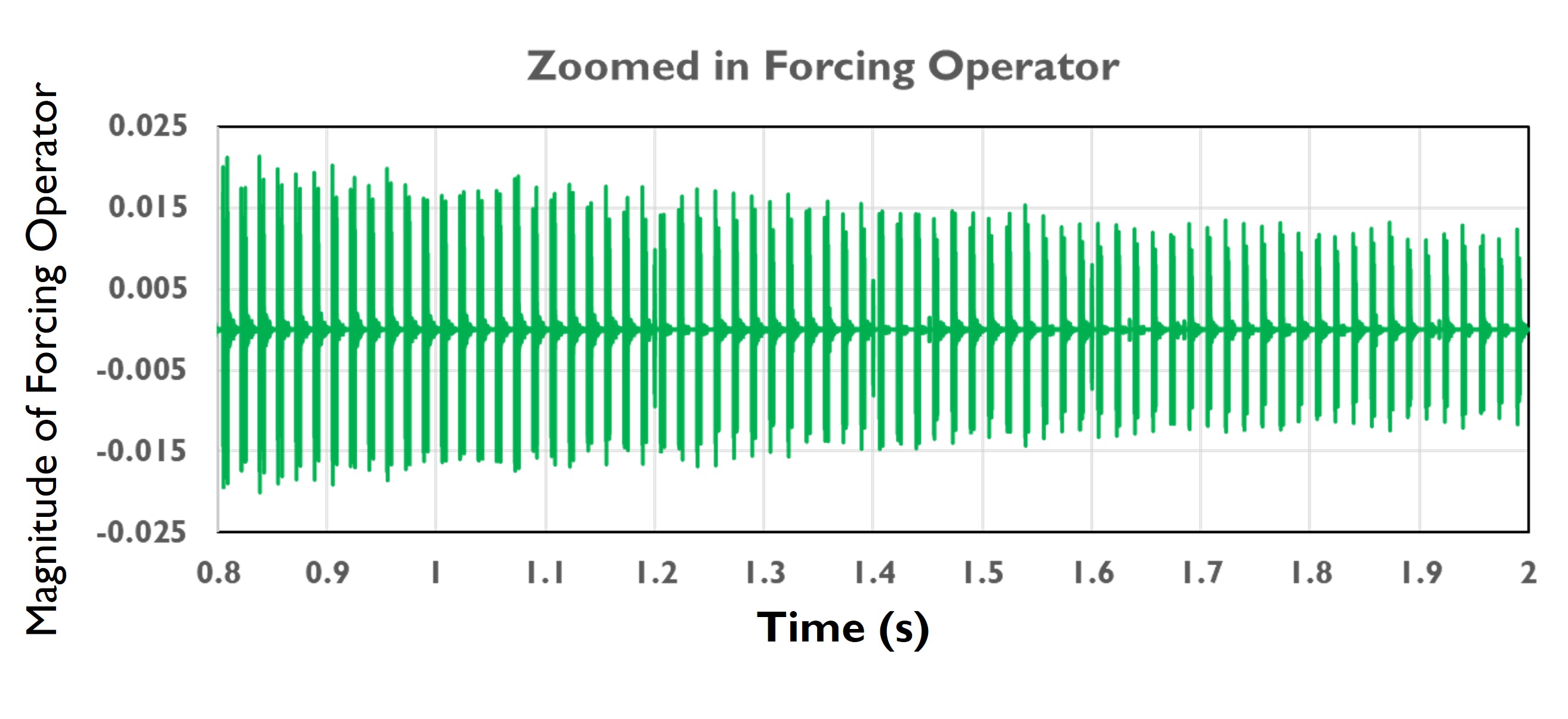}
  \caption{Zoomed view of forcing operator magnitude for Case A.}
  \label{fig:zoomed_forcing_A}
\end{figure}

\begin{figure}[htbp]
  \centering
  \includegraphics[width=3.5in]{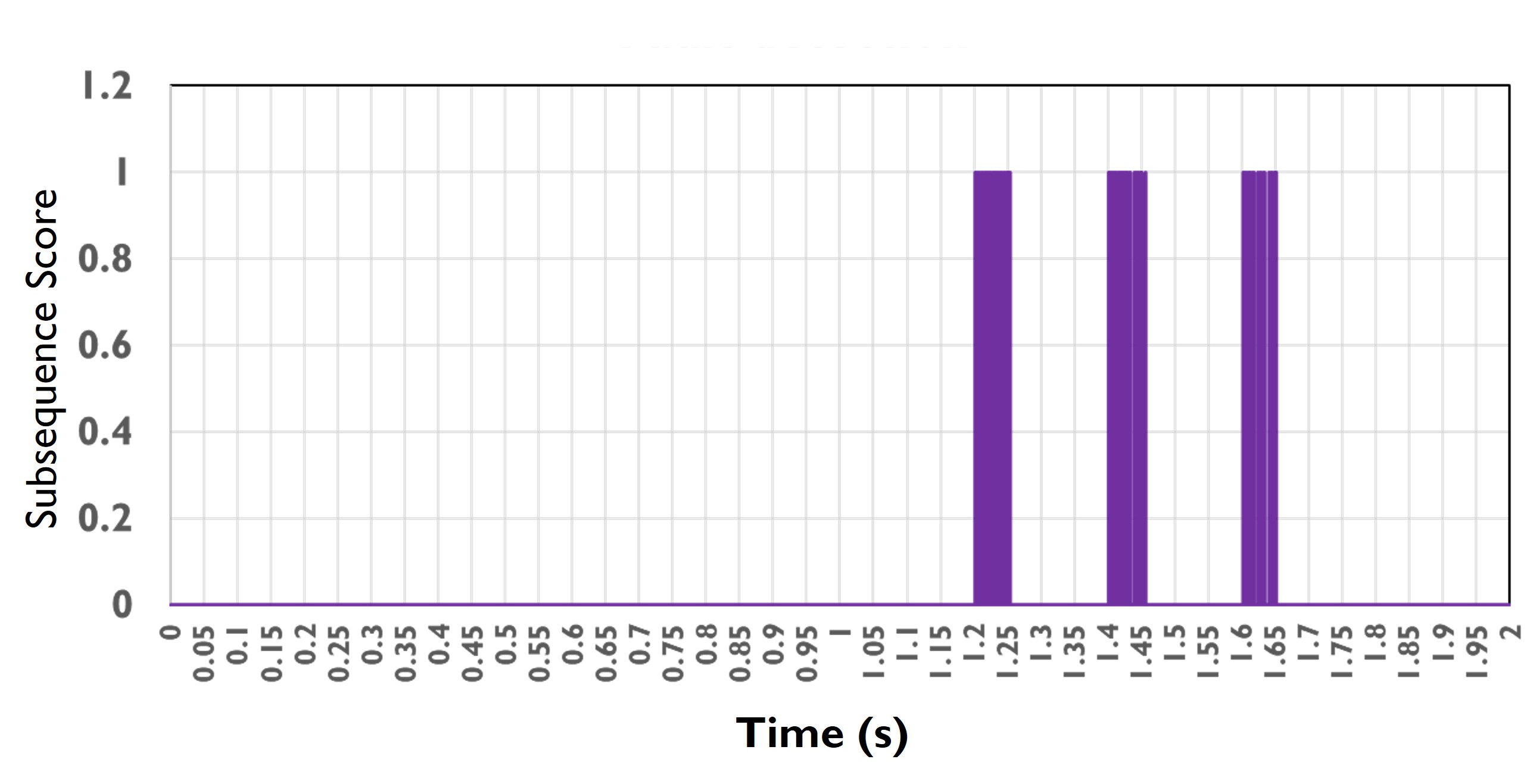}
  \caption{Subsequence scores for Case A.}
  \label{fig:score_A}
\end{figure}

\begin{figure}[htbp]
  \centering
  \includegraphics[width=3.5in]{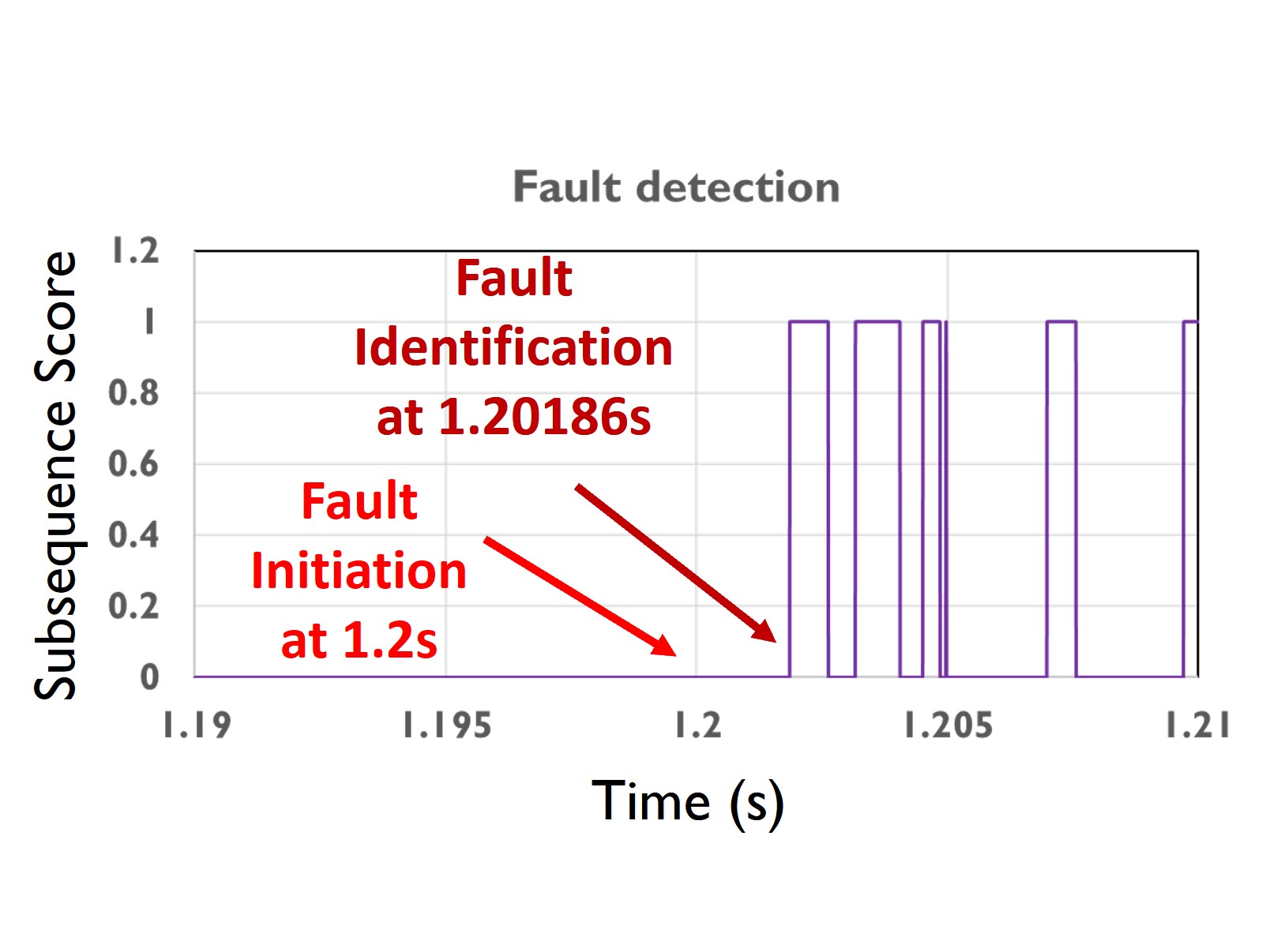}
  \caption{Zoomed view of subsequence scores for Case A.}
  \label{fig:zoomed_score_A}
\end{figure}

This primary current data is then processed using the Hankel alternative view of Koopman operator model to compute the forcing operator. Figures~\ref{fig:forcing_A} and~\ref{fig:zoomed_forcing_A} display the magnitude of the forcing operator over time. Figure~\ref{fig:forcing_A} shows the overall trend of the forcing operator magnitude, while Figure~\ref{fig:zoomed_forcing_A} provides a more detailed view of its behavior. It is observed that the forcing operator exhibits lower magnitudes for the duration of the HIF, contrasting with its behavior during normal operating conditions. This distinction is crucial for fault detection, as it transforms the subtle fault signatures in the current measurements into more pronounced variations in the forcing operator domain.
To further enhance fault detection capability, the Series2Graph algorithm is applied to the forcing operator values. This step yields the subsequence scores shown in Figures~\ref{fig:score_A} and~\ref{fig:zoomed_score_A}. 

Figure~\ref{fig:score_A} presents the overall pattern of subsequence scores throughout the simulation period, while Figure~\ref{fig:zoomed_score_A} offers a zoomed view that clearly highlights the anomalies at specific time instants. These anomalies, represented by sharp spikes in the subsequence scores, correspond to the occurrences of high-impedance arcing faults. The clear separation between the fault-induced spikes and the baseline scores during normal operation demonstrates the method's ability to effectively distinguish fault conditions from regular system behavior.

This multi-stage approach, combining the Hankel alternative view of Koopman operator for feature extraction, the forcing operator for signal transformation, and the Series2Graph algorithm for anomaly detection, proves highly effective in identifying high-impedance arcing faults that are otherwise difficult to detect from primary current measurements alone. The method's ability to clearly indicate anomalies at specific time instants, as shown in the subsequence score plots, provides a robust foundation for real-time fault detection and localization in complex distribution systems.

\begin{figure}[htbp]
  \centering
  \includegraphics[width=3.5in]{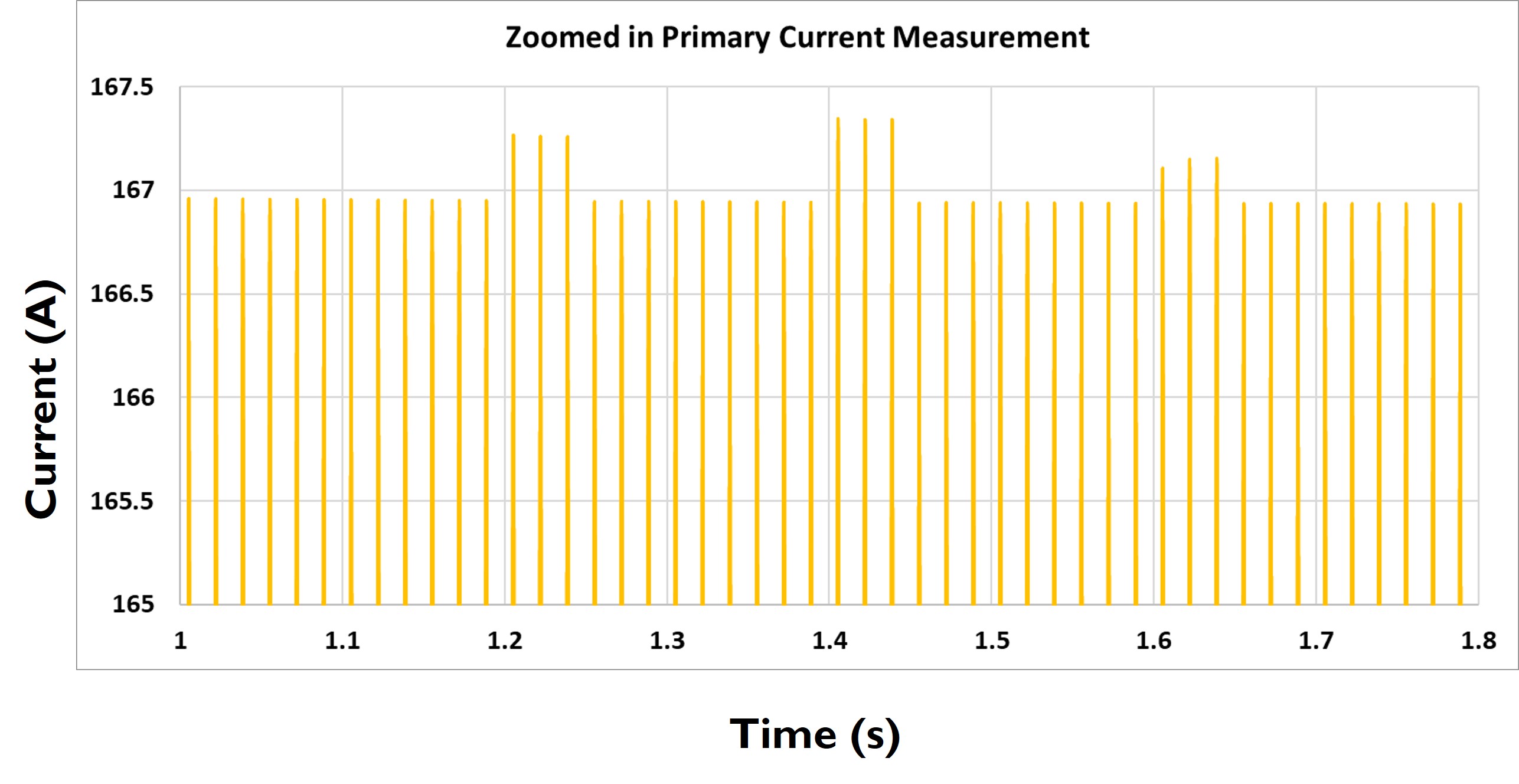}
  \caption{Zoomed view of primary current measurement for Case B.}
  \label{fig:zoomed_current_B}
\end{figure}

\begin{figure}[htbp]
  \centering
  \includegraphics[width=3.5in]{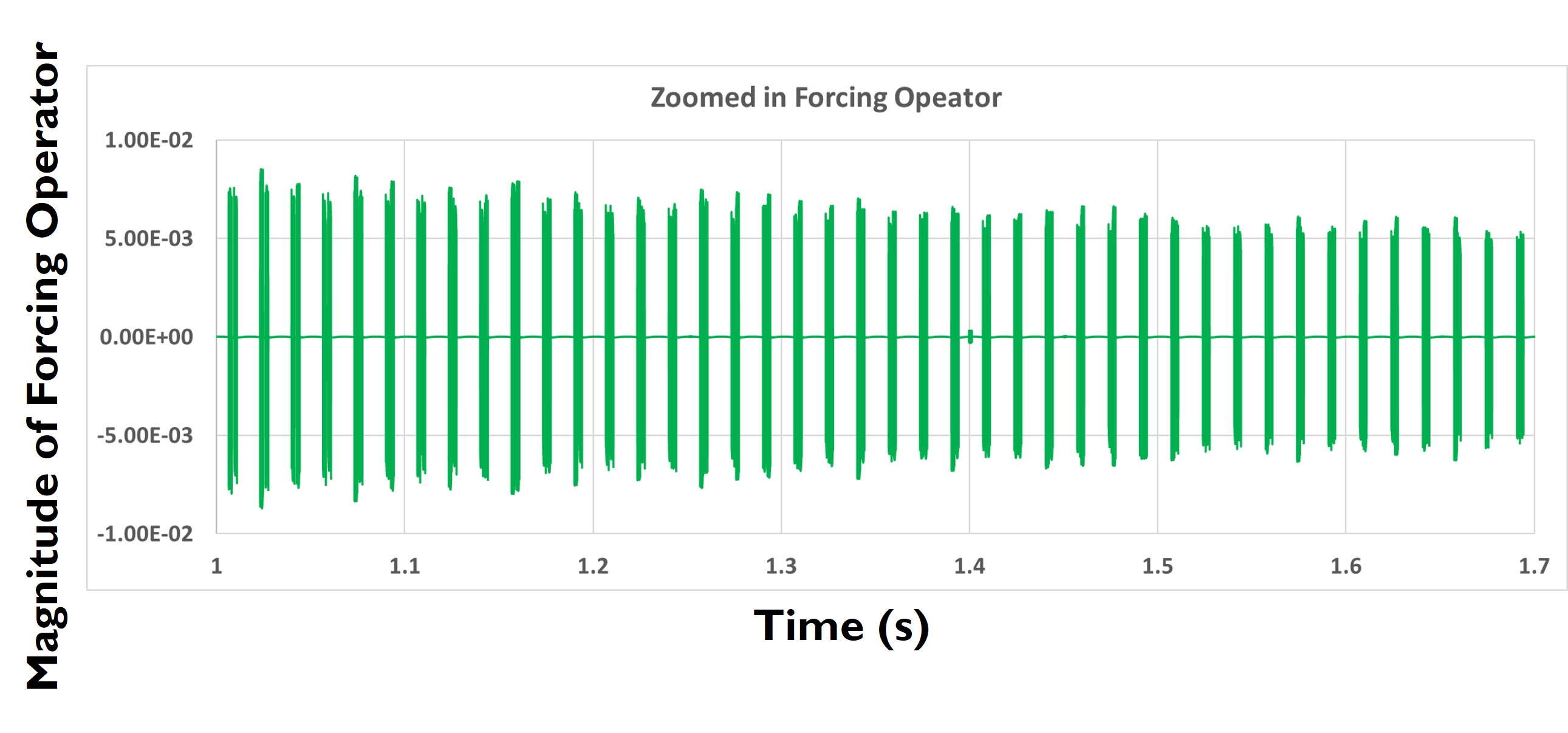}
  \caption{Zoomed view of forcing operator magnitude for Case B.}
  \label{fig:zoomed_forcing_B}
\end{figure}

\begin{figure}[htbp]
  \centering
  \includegraphics[width=3.5in]{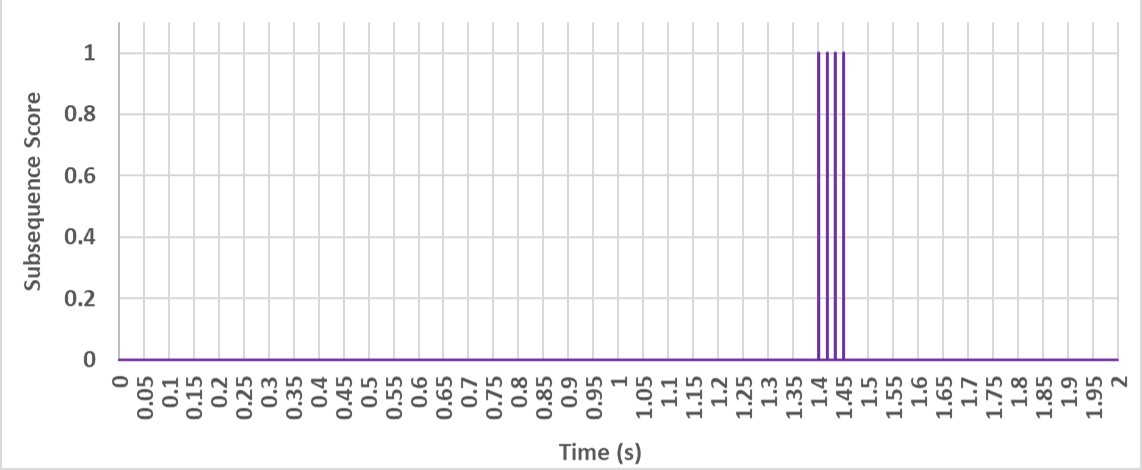}
  \caption{Subsequence scores for Case B.}
  \label{fig:score_B}
\end{figure}

\begin{figure}[htbp]
  \centering
  \includegraphics[width=3.5in]{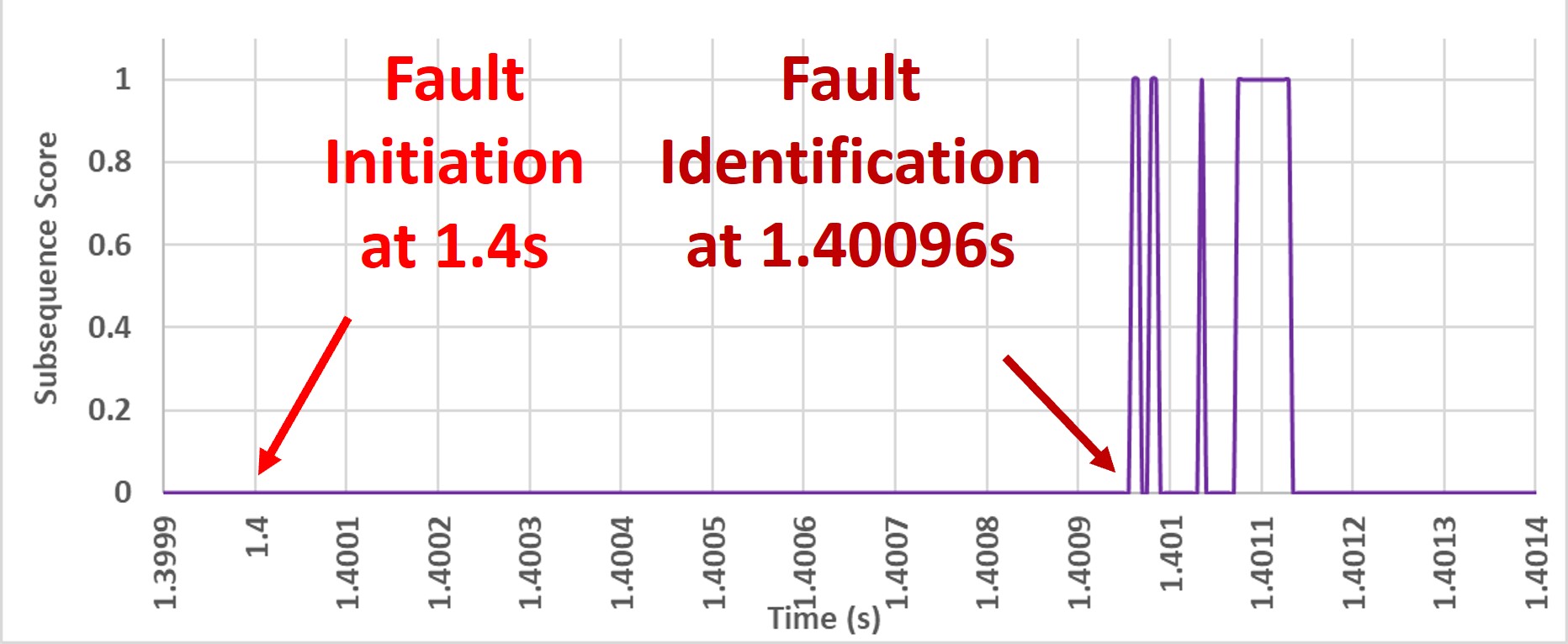}
  \caption{Zoomed view of subsequence scores for Case B.}
  \label{fig:zoomed_score_B}
\end{figure}

For Case B, which focuses on the secondary substation transformer, we observe distinct patterns in the measurements and analysis results. Figure~\ref{fig:zoomed_current_B} shows the zoomed-in primary current measurement at the secondary substation transformer. This detailed view reveals subtle current variations that may indicate the presence of high-impedance arcing faults, although these variations are not immediately distinguishable from normal load fluctuations.

Figure~\ref{fig:zoomed_forcing_B} displays the corresponding forcing operator magnitude for Case B. The forcing operator transformation accentuates the fault-related features in the signal, making the detection process more robust. The lower magnitudes observed during fault periods, similar to Case A, provide a crucial indicator for fault detection.

The subsequence scores, derived from the Series2Graph algorithm, are presented in Figures~\ref{fig:score_B} and~\ref{fig:zoomed_score_B}. Figure~\ref{fig:score_B} shows the overall trend of the scores throughout the simulation period, while Figure~\ref{fig:zoomed_score_B} offers a magnified view of the score variations. These scores exhibit distinct spikes corresponding to the simulated fault occurrences, clearly differentiating fault conditions from normal operations.
Our method successfully detects the high-impedance arcing faults at the simulated occurrences for both Case A and Case B, demonstrating its versatility across different system configurations. The detection performance is particularly noteworthy:
\begin{itemize}
\item In Case A, detection occurs within approximately 0.00186 seconds (about 1/10 of a cycle) after fault inception. 
\item For Case B, the fault is identified at 1.40096 seconds, representing an even faster detection time of about 1/16 of a power system cycle. 
\end{itemize}

It is important to note that the method effectively distinguishes between high-impedance arcing faults and normal operations, such as load switching and motor starting, even when these events have similar magnitudes in the current measurements. This discrimination ability is a significant advantage of our proposed approach, as it substantially reduces the likelihood of false positives in fault detection systems, a common challenge in traditional protection schemes.
The results demonstrate the robustness of the combined Koopman operator and Series2Graph approach in detecting and localizing high-impedance arcing faults across different system configurations. The high sample rate used in our simulations (over 2000 samples per cycle) contributes significantly to the method's precision. While this high sampling rate may present implementation challenges in some real-world systems, it represents the growing trend towards high-fidelity measurements in modern power systems.

The performance of the proposed method in both primary and secondary substation scenarios, coupled with its ability to differentiate between faults and normal transients, presents an effective solution for power system protection against arc faults. This approach offers improved reliability and safety for electrical distribution networks, potentially reducing outage times, minimizing equipment damage, and enhancing overall system resilience.

\section{Conclusion}
\label{section:Conclusion}

This study presents a comprehensive solution for detecting high-impedance arcing fault events in electrical distribution systems, addressing a critical challenge in power system protection. By leveraging real-time data from measuring devices at the primary side of substation transformers, our method successfully detects faults occurring at the secondary side without misoperating during similar characteristic events. The proposed approach combines the Koopman operator theory with advanced graph-based anomaly detection. First, we compute the Forcing Operator using Singular Value Decomposition of the Hankel Matrix constructed from real-time current measurements. This Forcing Operator is then analyzed using the Series2Graph algorithm, which effectively distinguishes fault conditions from normal operations.
The results demonstrate rapid and accurate fault detection, with identification times as low as 1/16 of a power system cycle. The method shows robust performance across different system configurations, successfully differentiating between high-impedance arcing faults and normal events such as load switching and motor starting. Future work could focus on optimizing the algorithm for real-time implementation and validating its performance under diverse fault scenarios.

\bibliographystyle{IEEEtran}
\bibliography{main}

\end{document}